\documentclass[11pt]{article}
\pdfoutput=1
\usepackage{sober}
\usepackage{amsthm,amsmath,fullpage,color,graphicx,subfigure,geometry}
\usepackage{hyperref,verbatim,algorithmic}
\usepackage[section,boxed]{algorithm}

\title{The Forgiving Tree: A Self-Healing Distributed Data Structure} 
\author{Tom Hayes\thanks{Toyota Technological Institute, Chicago, IL 60637;
email: {\tt hayest@tti-c.org}} 
\and Navin Rustagi \thanks{Department of Computer Science,  
University of New Mexico,  Albuquerque,  NM 87131-1386;
email: {\tt \{ navin, saia, amitabh\}@cs.unm.edu}. 
This research was partially supported by NSF CAREER Award 0644058,
NSF CCR-0313160, and an AFOSR MURI grant.} 
\and Jared Saia  \footnotemark[2] 
\and Amitabh Trehan  \footnotemark[2] }

\newtheorem{theorem}{Theorem}

\DeclareGraphicsExtensions{.pdf}

\newcommand{\RT}{\mathrm{RT}}
\newcommand{\ID}{\mathrm{ID}}
\newcommand{\SubRT}{\mathrm{SubRT}}

\newcommand{\will}{\mathrm{will}}
\newcommand{\heir}{\mathrm{heir}}
\newcommand{\helper}{\mathrm{helper}}
\newcommand{\parent}{\mathrm{parent}}
\newcommand{\hparent}{\mathrm{hparent}}

\newcommand{\children}{\mathrm{children}}
\newcommand{\hchildren}{\mathrm{hchildren}}
\newcommand{\isreadyheir}{\mathrm{isreadyheir}}
\newcommand{\degree}{\mathrm{degree}}
\newcommand{\diam}{\mathrm{diam}}
\newcommand{\ishelper}{\mathrm{ishelper}}
\newcommand{\nexthparent}{\mathrm{nexthparent}}
\newcommand{\nexthchildren}{\mathrm{nexthchildren}}
\newcommand{\nextparent}{\mathrm{nextparent}}

\newcommand{\Empty}{\emph{EMPTY}}

\renewcommand{\algorithmiccomment}[1]{// #1}

\begin{document}
\date{}
\maketitle

\thispagestyle{empty}

\begin{abstract}
We consider the problem of self-healing in peer-to-peer networks that
are under repeated attack by an omniscient adversary.  We assume that
the following process continues for up to $n$ rounds where $n$ is the
total number of nodes initially in the network: the adversary deletes
an arbitrary node from the network, then the network responds by
quickly adding a small number of new edges.

We present a distributed data structure that ensures two key
properties.  First, the diameter of the network is never more than
$O(\log \Delta)$ times its original diameter, where $\Delta$ is the
maximum degree of the network initially.  We note that for many
peer-to-peer systems, $\Delta$ is polylogarithmic, so the diameter
increase would be a $O(\log \log n)$ multiplicative factor.  Second,
the degree of any node never increases by more than $3$ over its
original degree.  Our data structure is fully distributed, has $O(1)$
latency per round and requires each node to send and receive $O(1)$
messages per round.  The data structure requires an initial setup
phase that has latency equal to the diameter of the original network,
and requires, with high probability, each node $v$ to send $O(\log n)$
messages along every edge incident to $v$.  Our approach is orthogonal
and complementary to traditional topology-based approaches to
defending against attack.
\end{abstract}

\section{Introduction}

Many modern networks are \emph{reconfigurable}, in the sense that the topology of the network can be changed by the
nodes in the network.  For example, peer-to-peer, wireless and mobile networks are reconfigurable.  More generally, many
social networks, such as a company's organizational chart; infrastructure networks, such as an airline's transportation
network; and biological networks, such as the human brain, are also reconfigurable.  Unfortunately, our mathematical and
algorithmic tools have not developed to the point that we are able to fully understand and exploit the flexibility of
reconfigurable networks.  For example, on August 15, 2007 the Skype network crashed for about $48$ hours, disrupting
service to approximately $200$ million users due to what the company described as failures in their ``self-healing
mechanisms''~\cite{garvey, fisher,malik, moore, ray, stone}.  We believe that this outage is indicative of a much broader
problem.

Modern reconfigurable networks have a complexity unprecedented in the history
of engineering: we are approaching scales of billions of components.
Such systems are less akin to a traditional engineering enterprise
such as a bridge, and more akin to a living organism in terms of
complexity.  A bridge must be designed so that key components never
fail, since there is no way for the bridge to automatically recover
from system failure.  In contrast, a living organism can not be
designed so that no component ever fails: there are simply too many
components.  For example, skin can be cut and still heal. Designing
skin that can heal is much more practical than designing skin that is
completely impervious to attack.  Unfortunately, current algorithms
ensure robustness in computer networks through hardening individual
components or, at best, adding lots of redundant components.  Such an
approach is increasingly unscalable.

In this paper, we focus on a new, \emph{responsive} approach for
maintaining robust reconfigurable networks.  Our approach is responsive in the sense
that it responds to an attack (or component failure) by changing the
topology of the network.  Our approach works irrespective of the
initial state of the network, and is thus orthogonal and complementary
to traditional non-responsive techniques.  There are many desirable
invariants to maintain in the face of an attack.  Here we focus only on
 the simplest and most fundamental invariants: ensuring the diameter of the network 
and the degrees of all nodes do not increase by much.

\medskip
\noindent {\bf Our Model:} We now describe our model of attack and
network response.  We assume that the network is initially a connected
graph over $n$ nodes.  An adversary repeatedly attacks the
network.  This adversary knows the network topology and our
algorithms, and it has the ability to delete arbitrary nodes from the
network.  However, we assume the adversary is constrained in that in
any time step it can only delete a single node from the network.  We
further assume that after the adversary deletes some node $x$ from the
network, that the neighbors of $x$ become aware of this deletion and
that the network has a small amount of time to react by adding and deleting some
edges.  This adversarial model captures what can happen when a
worm or software error propagates through the population of nodes.
Such an attack may occur too quickly for human intervention or for the
network to recover via new nodes joining.  Instead the nodes that
remain in the network must somehow reconnect to ensure that the
network remains functional.

We assume that the edges that are added can be added anywhere in the
network.  We assume that there is very limited time to react to
deletion of $x$ before the adversary deletes another node.  Thus, the
algorithm for deciding which edges to add between the neighbors of $x$
must be fast.

\medskip
\noindent {\bf Our Results:}  A naive approach to this problem is
simply to 'surrogate' one neighbor of the deleted node to take on the
role of the deleted node, reconnecting the other neighbors to this
surrogate. However, an intelligent adversary can always cause this approach to 
increase the degree of some node by $\theta(n)$. On the other hand, we may try to keep the degree
increase low by connecting neighbors of the deleted node as a straight
line, or by connecting the neighbors of the deleted node in a binary tree.  However, for both of these techniques the diameter can increase by $\theta(n)$ over multiple deletions by an intelligent adversary~\cite{BomanSAS06, SaiaTrehanIPDPS08}.

In this paper, we describe a new, light-weight distributed data structure that ensures that: 1) the diameter of the network never increases by more than $\log \Delta$ times its original diameter, where $\Delta$ is the
maximum degree of a node in the original network; and 2) the degree of
any node never increases by more than $3$ over over its original
degree.  Our algorithm is fully distributed, has $O(1)$ latency per
round and requires each node to send and receive $O(1)$ messages per
round.   The formal statement and proof of these results is in Section~\ref{subsec: upperbounds}.  Moreover, we show (in Section~\ref{subsec:
lowerbounds}) that
in a sense our algorithm is asymptotically optimal, since any algorithm that increases node degrees by no more than a
constant must, in some cases, cause the diameter of a graph to increase by a $\log \Delta$ factor.

The algorithm requires a one-time setup phase  to do the
following two tasks.  First, we must find a breadth first spanning
tree of the original network rooted at an arbitrary node.  This can be
done with latency equal to the diameter of the original network, and,
with high probability, each node $v$ sending $O(\log n)$ messages
along every edge incident to $v$ as in the algorithm due to
Cohen~\cite{Cohen}. The second task required is to set up a simple
data structure for each node that we refer to as a will.  This will,
which we will describe in detail in the Section~\ref{sec:algorithm}, gives
instructions for each node $v$ on how the children of $v$ should
reestablish connectivity if $v$ is deleted.  Creating the will
requires $O(1)$ messages to be sent along the parent and children
edges of the global breadth-first search tree created in the first
task.
 

\medskip
\noindent {\bf Related Work:} There have been numerous papers that
discuss strategies for adding additional capacity and rerouting in
anticipation of failures \cite{ doverspike94capacity,
frisanco97capacity, iraschko98capacity, murakami97comparative,
caenegem97capacity, xiong99restore}.  Here we focus on results that
are responsive in some sense.  M\'{e}dard, Finn, Barry, and Gallager
\cite{medard99redundant} propose constructing redundant trees to make
backup routes possible when an edge or node is deleted.  Anderson,
Balakrishnan, Kaashoek, and Morris \cite{anderson01RON} modify some
existing nodes to be RON (Resilient Overlay Network) nodes to detect
failures and reroute accordingly. Some networks have enough redundancy
built in so that separate parts of the network can function on their
own in case of an attack~\cite{goel04resilient}.  In all these past
results, the network topology is fixed.  In contrast, our algorithm
adds edges to the network as node failures occur.  Further, our
algorithm does not dictate routing paths or specifically require
redundant components to be placed in the network initially.  In this
paper, we build on earlier work in~\cite{BomanSAS06, SaiaTrehanIPDPS08}.


There has also been recent research in the physics community on
preventing cascading failures.  In the model used for these results,
each vertex in the network starts with a fixed capacity. When a vertex
is deleted, some of its ``load'' (typically defined as the number of
shortest paths that go through the vertex) is diverted to the
remaining vertices.  The remaining vertices, in turn, can fail if the
extra load exceeds their capacities. Motter, Lai, Holme, and Kim have
shown empirically that even a single node deletion can cause a
constant fraction of the nodes to fail in a power-law network due to
cascading failures\cite{holme-2002-65, motter-2002-66}. Motter and Lai
propose a strategy for addressing this problem by intentional removal
of certain nodes in the network after a failure begins
\cite{motter-2004-93}.  Hayashi and Miyazaki propose another strategy,
called emergent rewirings, that adds edges to the network after a
failure begins to prevent the failure from
cascading\cite{hayashi2005}.  Both of these approaches are
shown to work well empirically on many networks.  However, unfortunately, they
perform very poorly under adversarial attack.

\section{Delete and Repair Model}
\label{sec:prelim}

We now describe the details of our delete and repair model.  Let $G = G_0$ be an arbitrary graph on $n$ nodes,
which represent processors in a distributed network.  One by one, the Adversary deletes nodes until none are left.  After each deletion, the Player gets to add some new edges to the graph, as well as deleting old ones.   The Player's goal is to maintain connectivity in the network, keeping the diameter of the graph small.  At the same time, the Player wants to minimize the resources spent on this task,
in the form of extra edges added to the graph, and also in 
terms of the number of connections maintained by each node
at any one time (the degree increase).  We seek an algorithm which gives performance guarantees 
under these metrics for each of the $n!$ possible deletion orders.

Unfortunately, the above model still does not capture the
behaviour we want, since it allows for a centralized Player
who ignores the structure of the original graph, and simply
installs and maintains a complete binary tree, using a leaf 
node to substitute for each deleted node. 

To avoid this sort of solution, we require a distributed
algorithm which can be run by a processor at each node.
Initially, each processor only knows its neighbors in $G_0$,
and is unaware of the structure of the rest of the $G_0$.
After each deletion (forming $H_t$), only the neighbors of the deleted
vertex are informed that the deletion has occurred.
After this, processors are allowed to communicate by
sending a limited number of messages to their direct 
neighbors.  We assume that these messages are always
sent and received successfully.  The processors may 
also request new edges be added to the graph to form $G_t$.
The only synchronicity assumption we make is that the
next vertex is not deleted until the end of this round
of computation and communication has concluded.
To make this assumption more reasonable, the per-node
communication should be $O(1)$ bits, and should 
moreover be parallelizable so that the entire protocol
can be completed in $O(1)$ time if we assume synchronous
communication.

We also allow a certain amount of pre-processing to be
done before the first deletion occurs.  This may, for
instance, be used by the processors to gather some
topological information about $G_0$, or perhaps to 
coordinate a strategy.  Another success metric is
the amount of computation and communication needed
during this preprocessing round.  Our full model is described as Model~\ref{algo:model-2} below.

\floatname{algorithm}{Model}

\begin{algorithm}[h!]
\caption{The Delete and Repair Model -- Distributed View.}
\label{algo:model-2}
\begin{algorithmic}
\STATE Each node of $G_0$ is a processor.  
\STATE Each processor starts with a list of its neighbors in $G_0$.
\STATE Pre-processing: Processors may send messages to and from
their neighbors.
\FOR {$t := 1$ to $n$}
\STATE Adversary deletes a node $v_t$ from $G_{t-1}$, forming $H_t$.
\STATE All neighbors of $v_t$ are informed of the deletion.
\STATE {\bf Recovery phase:}
\STATE Nodes of $H_t$ may communicate (asynchronously, in parallel) 
with their immediate neighbors.  These messages are never lost or
corrupted, and may contain the names of other vertices.
\STATE During this phase, each node may insert edges
joining it to any other nodes as desired. 
Nodes may also drop edges from previous rounds if no longer required.
\STATE At the end of this phase, we call the graph $G_t$.
\ENDFOR
\STATE {\bf Success metrics:} Minimize the following ``complexity'' measures:
 \begin{enumerate}
\item{\bf Degree increase.} $\max_{t<n} \max_{v} \degree(v,G_t) - \degree(v,G_0)$
\item {\bf Diameter stretch.} $\max_{t<n} \diam(G_t) / \diam(G_0)$
\item{\bf Communication per node.} The maximum number of bits sent by a single node in a single recovery round.
\item{\bf Recovery time.} The maximum total time for a recover round,
assuming it takes $1$ bit no more than $1$ time unit to traverse any edge and unlimited local computational power at each node.
\end{enumerate}
\end{algorithmic}
\end{algorithm}

\section{The Forgiving Tree algorithm}
\label{sec:algorithm}

At a high level, our algorithm works as follows. We begin with a
rooted spanning tree $T$, which without loss of generality may as well
be the entire network.

\begin{figure}[t!]
\centering
\includegraphics[scale=0.4]{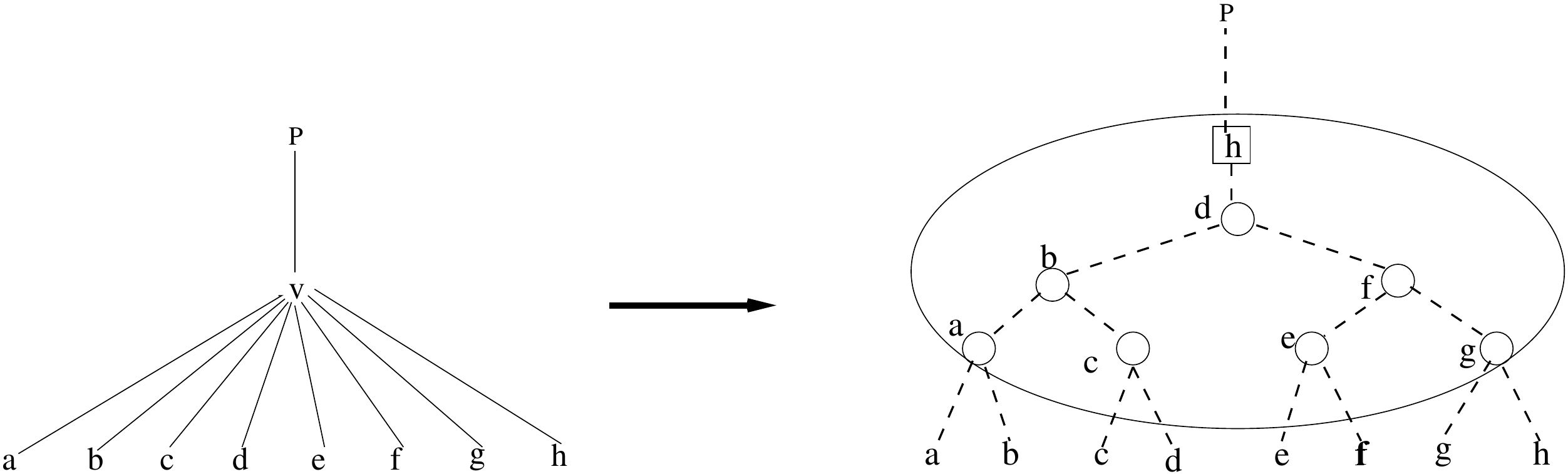}
\caption{Deleted node $v$ replaced by its Reconstruction Tree. The nodes in the oval are helper nodes. Regular helper
nodes are depicted by circles and the heir helper node by a rectangle.}
 \label{fig: RT}
\end{figure}

Each time a non-leaf node $v$ is deleted, we think of it as being
replaced by a balanced binary tree of ``virtual nodes,'' 
with the leaves of the virtual tree taking $v$'s
place as the parents of $v$'s children. Depending on certain
conditions explained later, the root of this ``virtual tree'' or
another virtual node (known as $v$'s \emph{heir}---this will be
discussed later) takes $v$'s place as the child of $v$'s
parent. This is illustrated in figure \ref{fig: RT}.
Note that each of the virtual nodes which was added is of
degree $3$, except the heir, if present.

When a leaf node is deleted,
we do not replace it.  However, if the parent of the deleted 
leaf node was a virtual node, its degree has now reduced
from $3$ to $2$, at which point we consider it redundant
and ``short-circuit'' it, removing it from the graph,
and connecting its surviving child directly to its parent.
This helps to ensure that, except for heirs, every virtual
node is of degree exactly $3$.

After a long sequence of such deletions, we are left with a
tree which a patchwork mix of virtual nodes and original nodes.
We note that the degrees of the original nodes never increase
during the above procedure.
Also, because the virtual trees are balanced binary trees, the deletion of
a node $v$ can, at worst, cause the distances between its
neighbors to increase from $2$ to $2 \lceil \log d  \rceil$, where 
$d$ is the degree of $v$.  This ensures that,
even after an arbitrary sequence of deletions, the distance
between any pair of surviving actual nodes has not increased
by more than a $\lceil \log \Delta \rceil$ factor, 
where $\Delta$ is the maximum degree of the original tree.

Since our algorithm is only allowed to add edges and not nodes, 
we cannot really add these virtual nodes to the network.
We get around this by assigning each virtual node to an actual
node, and adding new edges between actual nodes in order to 
allow ``simulation'' of each virtual node.  More precisely,
our actual graph is the homomorphic image of the tree
described above, under a graph homomorphism which fixes 
the actual nodes in the tree and maps each virtual node
to a distinct actual node which is ``simulating'' it.
The existence of such a mapping is a consequence of the 
fact that all the virtual nodes have degree $3$, except heirs,
which have degree $2$ (and there are not too many of these), 
and will be proved later.
Note that, because each actual node ever simulates one
virtual node at a time, and virtual nodes have degree at most $3$,
this ensures that the maximum degree increase of our algorithm
is at most $3$.

The heart of our algorithm is a very efficient distributed algorithm 
for keeping track of which actual node is assigned to simulate
each virtual node, so that the replacement of each deleted node by
its virtual tree can be done in $O(1)$ time.  
We accomplish this using a system of ``wills,'' 
in which each vertex $v$ instructs each of its children (or their ``heirs'') 
in the event of $v$'s deletion, how to simulate the
virtual tree replacing $v$, and also the virtual node $v$ was
simulating (if any).

\begin{figure}[h!]
\centering
\includegraphics[scale=0.4] {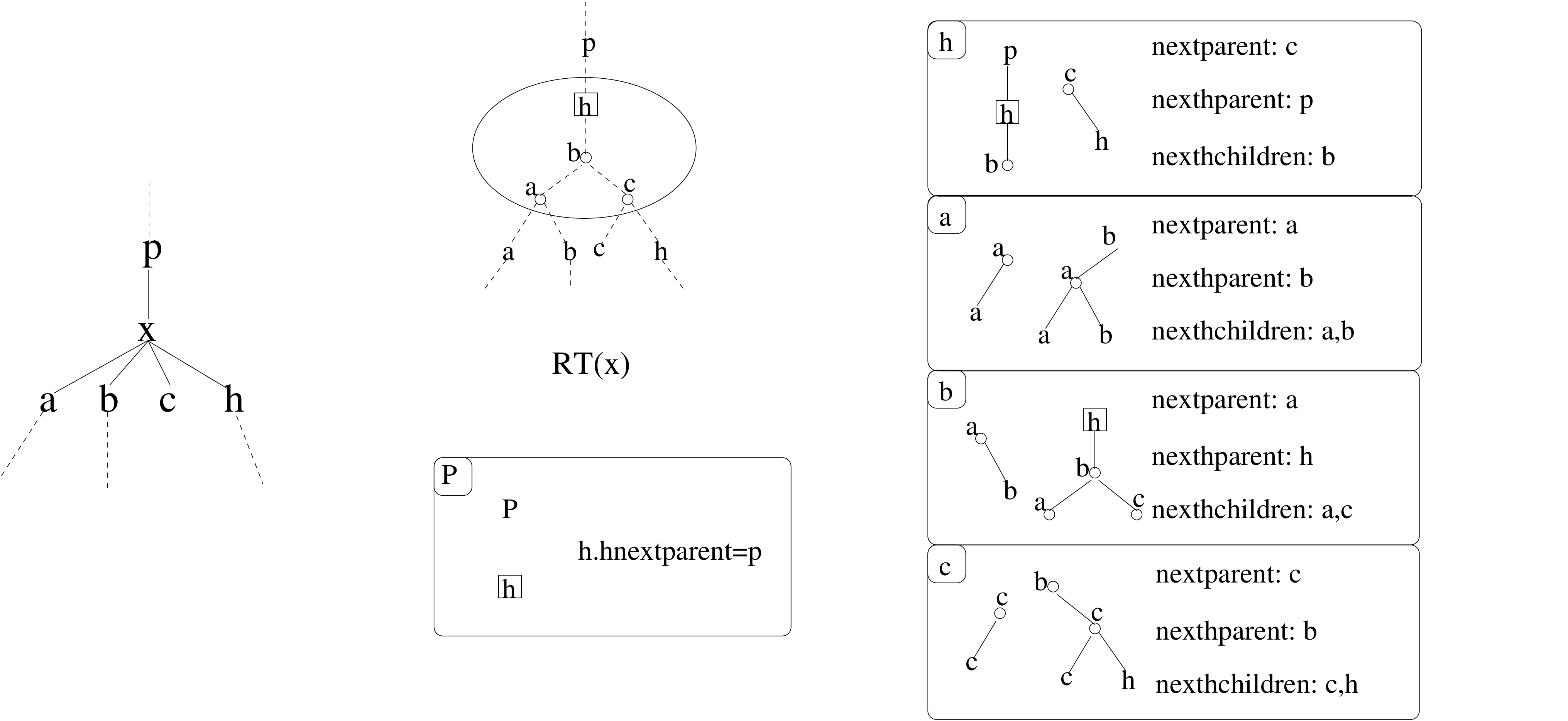}
\caption{The leftmost column shows a small segment of the network.  The RT(x) corresponding to this figure is shown. 
Every neighbor of node $x$ stores the portion of $\RT(x)$ relevant to it. Each rectangular box is labelled with a neighbor and shows the portions and the value of the corresponding fields . }
\label{fig: RTbreakup}
\end{figure}

This will is prepared in advance, before $v$'s deletion, and entrusted
to $v$'s children or their surviving heirs.  An example of this is
shown in figure \ref{fig: RTbreakup}.  Certain events, such as the
deletion of one of $v$'s children, or a change in which virtual node
$v$ is simulating, may cause $v$ to revise its will, informing the
affected children or their surviving heirs.  As shall be seen, the
total size and number of messages which must be sent is $O(1)$ per
deleted vertex. In addition, there is a startup cost for communicating
the initial wills; this is $O(1)$ per edge in the original network.

\subsection{Detailed description} 

To begin with, in Table~\ref{tab: nodedata} we list the data kept by
each real node $v$ required for the ForgivingTree algorithm. We have
four main classes of fields, according to the way they are used by the
node.  `Current fields' give a nodes present configuration and status
in the tree. `Reconstruction fields' hold the data needed for a node
to reconstruct connections when one of its neighbors gets
deleted. `Helper fields' hold information with regard to the helper
node being simulated by this node. Each node also stores some special
flags with regard to its helper or heir status. In the description
that follows, we shall refer directly to these fields.

\begin{table}[!h]
\begin{tabular}{|l|p{4.5in}|}
\hline 
\textbf{Current fields}& Fields having  information about a node's current neighbors.\\ \cline{2-2}
  \texttt{parent(v)}& Parent of $v$.\\
  \texttt{children(v)}& Children of $v$.\\
  \texttt{SubRT(v)}& Stores the Reconstruction Tree ($\RT$) of $v$ minus a possible helper node
simulated by $\heir(v)$.  This tree of helper and real nodes that shall replace $v$ if $v$ is deleted.\\
 \texttt{heir(v)} & The heir of $v$.\\
\hline
\textbf{Helper fields} & Fields specifying a node's role as a helper node.\\ \cline{2-2}
 \texttt{hparent(v)}& Parent of the helper node  $v$ may be simulating. \\
 \texttt{hchildren(v)}& Children of the helper node $v$ may be simulating.\\
\hline
\textbf{Reconstruction fields}& Fields used by a node to reconstruct its connections when its neighbor is deleted.\\ \cline{2-2}
 \texttt{nextparent(v)}& The node which will be the next $\parent$ of $v$. \\
 \texttt{nexthparent(v)}& The node which will be the next $\hparent$ of $v$. \\
 \texttt{nexthchildren(v)}& The node(s) which will be the next $\hchildren$ of $v$.\\
\hline
\textbf{Flags}& Specifying a node's helper or heir status.\\ \cline{2-2}
\texttt{ishelper(v)}& (boolean field). True if $v$ is simulating a \emph{helper} node, false otherwise.\\
\texttt{isreadyheir(v)}&  (boolean field). True if $v$ is simulating an heir in ready state, false otherwise (wait or deployed state). \\
\hline
\end{tabular}
\caption{The fields maintained by a node $v$}
\label{tab: nodedata}
\end{table}

%
%

At the top level, our algorithm is specified as Algorithm \ref{algo: forgiving}~: \textsc{Forgiving tree}. Algorithm \ref{algo: forgiving}~ uses Algorithms 2 to 9, which will be described at the appropriate places. As referred to earlier, \textsc{Forgiving tree} works on a tree which may be obtained from the original graph during a preprocessing phase. The next stage is an initialisation phase in which the appropriate data structures are setup. Once these are setup, the network is ready to face the adversarial attacks as and when they happen. 
  
\subsubsection{The Initialization phase}
This phase is specified in Algorithm \ref{algo: init}~: \textsc{Init()}. We assume each node $v$ has a unique
identification number  which we call $ID(v)$. Every node in the tree initializes the fields we have listed in Table~\ref{tab: nodedata}.   In our descriptions if no data is available or appropriate for a field, we set it to $\Empty$.
Since no deletion has happened yet and there are no helper nodes in the system, the helper fields are set to $\Empty$.
The current fields $\parent(v)$ and $\children(v)$ are assigned pointers to the parent and children of $v$. Of course,
if $v$ is a leaf node $\children(v)$ is $\Empty$ and if $v$ is the root of the tree $\parent(v)$ is $\Empty$. 

 As stated earlier, the heart of our algorithm is the system of
$\will$s created by nodes and distributed among its neighbors. 
The will of a node, $v$, has two parts: firstly, a Reconstruction Tree
($\SubRT(v)$), which will replace $v$ when 
it is deleted by the Adversary, 
and secondly, the delegation of $v$'s helper responsibilities 
(if any) to a child node, $\heir(v)$.
For concreteness, we initially designate the child of $v$ with
the highest $\ID$ as $\heir(v)$.  In the event that $\heir(v)$
is deleted, its role will be taken over by its heir, if any.
If $\heir(v)$ is a leaf when it is deleted, then $v$ will 
designate its new heir to be the surviving child whose helper
node has just decreased in degree from $3$ to $2$.

Algorithm \ref{algo: genRT}: \textsc{GenerateSubRT} computes
$\SubRT(v)$. If the node $v$ has no helper responsibilities, as is
during this phase, $\RT(v)$ is simply $\SubRT(v)$ with a helper node
simulated by $\heir(v)$ appended on as the parent of the root of
$\SubRT(v)$. Figure \ref{fig: RT} and Turn 1 in Fig~\ref{fig: story}
depict such Reconstruction Trees. If the node $v$ has helper
responsibilities $\RT(v)$ is the same as $\SubRT(v)$. Node $v$ uses
Algorithm \ref{algo: genRT} to compute $\SubRT(v)$ as follows: All the
children of $v$ are arranged as a single layer in sorted (say,
ascending) order of their $\ID$s. Then a set of helper nodes - one
node for each of the children of $v$ except the heir are arranged
above this layer so as to construct a balanced binary search tree
ordered on their $ID$s.

 The last step of the initialization process is to finalize the will and transmit it to the children. Each child is
given only the portion of the will relevant to it. Thus, only this portion needs to be updated whenever a will changes.
The division of $\RT$ into these portions is shown in figure \ref{fig: RTbreakup}.  There are fundamentally two
different kinds of wills : one prepared by leaf nodes who have helper responsibilities and the other by non-leaf nodes.
Obviously, during the initialisation phase, only the second kind of will is needed. This is finalized and distributed as
shown in Algorithm \ref{algo: makewill}: \textsc{MakeWill}. 
The children of $v$ initialize their reconstruction fields
with the values from $\SubRT(v)$. If later $v$ gets deleted these values will be copied to present and helper fields
such that $\RT(v)$ is instantiated. Notice that the role the heir will assume is decided according to whether $v$ is a
helper node or not. Since $v$ cannot be a helper node in this phase, the heir node simply sets its reconstruction
fields so as to be between the root of $\SubRT(v)$ and $\parent(v)$. In this case when $\RT(v)$ will be instantiated,
the helper node simulated by $\heir(v)$ shall have only one child: we will say that $\heir(v)$ is in the ready phase
(explained later) and set the flag $\isreadyheir(v)$ to true. In the initialization phase both the $\isreadyheir$ and 
$\ishelper $ flags will be set to false. 

This completes the setup and initialization of the data structure. Now our network is ready to handle adversarial
attacks. In the context of our algorithm, there are two main events that can happen repeatedly and  need to be handled
differently:
\subsubsection{Deletion of an internal node}
 The healing that happens on deletion of a non-leaf node is specified in Algorithm \ref{algo: fixnode}:
\textsc{FixNodeDeletion}. In our model, we assume that the failure of a node is only detected by its neighbors in the
tree, and it is these nodes which will carry out the healing process and update the changes wherever required. If the 
node $v$ was deleted, the first step in the reconstruction process is to put $\RT$ into place according to Algorithm
\ref{algo: makeRT}: \textsc{makeRT}. Note that all children of $v$ have lost their parent. Let us discuss the
reconstruction performed by non-heir nodes first. They make an edge to their new parent (pointer to which was
available as nextparent()) and set their current fields. Then they take the role of the helper nodes as specified in
$\RT(v)$ and Algorithm \ref{algo: makehelper}: \textsc{MakeHelper} and make the required edges and field changes to
instantiate $\RT(v)$.

To understand what the heir node does in this case,  it will be useful here to have a small discussion on the states of
a regular/heir node:\\

\begin{figure}[h!]
\centering
\includegraphics[scale=0.8] {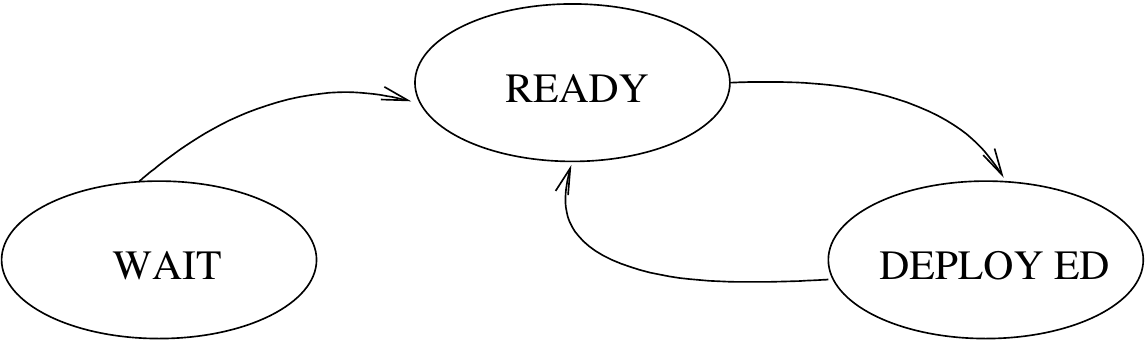}
\caption{The states of a node with respect to helper duties: Waiting, Ready and Deployed}
\label{fig: nodestates}
\end{figure}
  
\noindent {\bf States of a heir/regular node:} Consider a node $v$ and its heir $h$. From the point of view of $h$, we
can imagine $h$  to be in one of three states which we  call \emph{wait}, \emph{ready} and \emph{deployed}. These states
are illustrated in figure ~\ref{fig: nodestates}. For ease of discussion, let us call the helper node that a node is simulating
$\helper(node)$.
In brief, a node is considered to be in the wait state when it has no helper
responsibilities, in the ready state when  $\helper(node)$ with one child, and in the deployed state when  $\helper(node)$
 has  two children (which is the maximum possible).  Notice that the node can  be in the wait state only when  $v$ has
not been deleted and thus, $h$ has assumed no helper responsibilities. It only has the will of $v$ and is in limbo
with regard to helper duties. Now consider the case when $v$ gets deleted.  Following are the possibilities:
\begin{itemize}
\item \emph{node $v$ had no helper responsibilities}: This happens when $v$'s original parent was not deleted. Thus,
 $v$ could be a regular child or a heir in the wait state. On $v$'s deletion $h$ moves to the ready state and sets its
flag $\isreadyheir$ to True. This is the state in which  $\helper(h)$ has only one child i.e.
the root of $\SubRT(v)$. This happens when $h$ executes its portion of the will of $v$ using Algorithm \ref{algo: makeRT}:
\textsc{makeRT}. Note that this may not be the final state for the helper node of $h$, and is thus called the ready
state.
\item \emph{node $v$ had helper responsibilities}: There are two further possibilities:
\begin{itemize}
\item \emph{$\helper(v)$ had one child}: This can only happen when $v$ was a heir node  in the ready state. Thus,
$v$'s flags $\ishelper$ and $\isreadyheir$ were both set to True. Node $h$
will take over the helper responsibilities of $v$ and thus, in turn, $h$ will now have one child i.e. will be in the
ready state and will set its flags $\ishelper$ and $\isreadyheir$ to True. Notice that if $v$ was an heir, $h$ will
now also take over those responsibilities, and on future deletions of $v$'s ancestors could move further up the tree
either as an heir in ready state or in a deployed state become a full helper node.
\item \emph{$\helper(v)$ had two children}: Node $v$ could be a regular child or heir. $h$ will fully take over the
helper responsibilities of $v$, and thus $\helper(h)$ shall acquire two children and move on to the deployed state.
Notice that previously $h$ could have been in either  wait or  ready state. Since  it is now not in the ready
state, it will set its $\isreadyheir$ flag to False and $\ishelper$ flag to True.
\end{itemize}
\end{itemize}
 It is easy to see that a regular i.e. non-heir node can be in either wait or deployed state.\\
 
 Here we also define the following operation, which is used in Algorithms  \ref{algo: fixleaf} and \ref{algo: makeRT}:
\begin{description}
\item[bypass(x):] \emph{Precondition}: $|\hchildren(x)| = 1$ i.e the helper node has a single child. \emph{Operation}:
\emph{Delete} $\helper(x)$ i.e. $\hparent(x)$ and $\hchildren(x)$  remove their edges with $x$ and make a
new edge between themselves.  $\hparent(x) \leftarrow \Empty$; $\hchildren(x) \leftarrow \Empty$.
\end{description}

  We can now easily see how the heir of $v$, $h$ takes part in the reconstruction according to Algorithm
\ref{algo: makeRT}: \textsc{makeRT}. Node $h$ can be either in wait state or ready state. If it is in the wait
state it simply takes its helper responsibilities according to Algorithm \ref{algo: makehelper}: \textsc{MakeHelper},
as in turn 1 of figure~\ref{fig: story} .
Note that here $h$ checks  if it has moved to the ready state and sets its $\isreadyheir$ flag accordingly.
 If $h$ was already in ready state, it relinquishes its present helper role and moves on to the new helper
role. To relinquish its present role, node $h$ intimates  $\hparent(h)$ and $\hchildren(h)$, and together they accomplish
this as specified by the operation bypass($h$). Turn 2 in Fig~\ref{fig: story} illustrates this. 

Once $\RT(v)$ is in a place, there may be a need for the parent of $v$ to recompute its will. This happens only when
$v$ did not already have a helper role or equivalently when $\heir(v)$ moves to a ready state. Lines \ref{algline:
ND2} to \ref{algline: ND6} of Algorithm \ref{algo: fixnode} deals with this situation. Node $\parent(v)$ simply
replaces $v$ by $\heir(v)$ in its will and retransmits it.  At the end of this healing process, the children of the
deleted nodes check if they need to leave the second kind of will, which we call a \emph{LeafWill}. This will is
required only for those nodes which are leaves in our tree and have virtual responsibilities. Since they have no
children to take over their helper responsibilities they leave this responsibility to their parent. We will discuss
this in greater detail in the next section.

\subsubsection{Deletion of a leaf node}
\begin{figure}[h!]
\centering
\subfigure[$\helper(v)$ is ancestor of $v$.]{\label{sfig: ldc1} \includegraphics[scale=0.3]{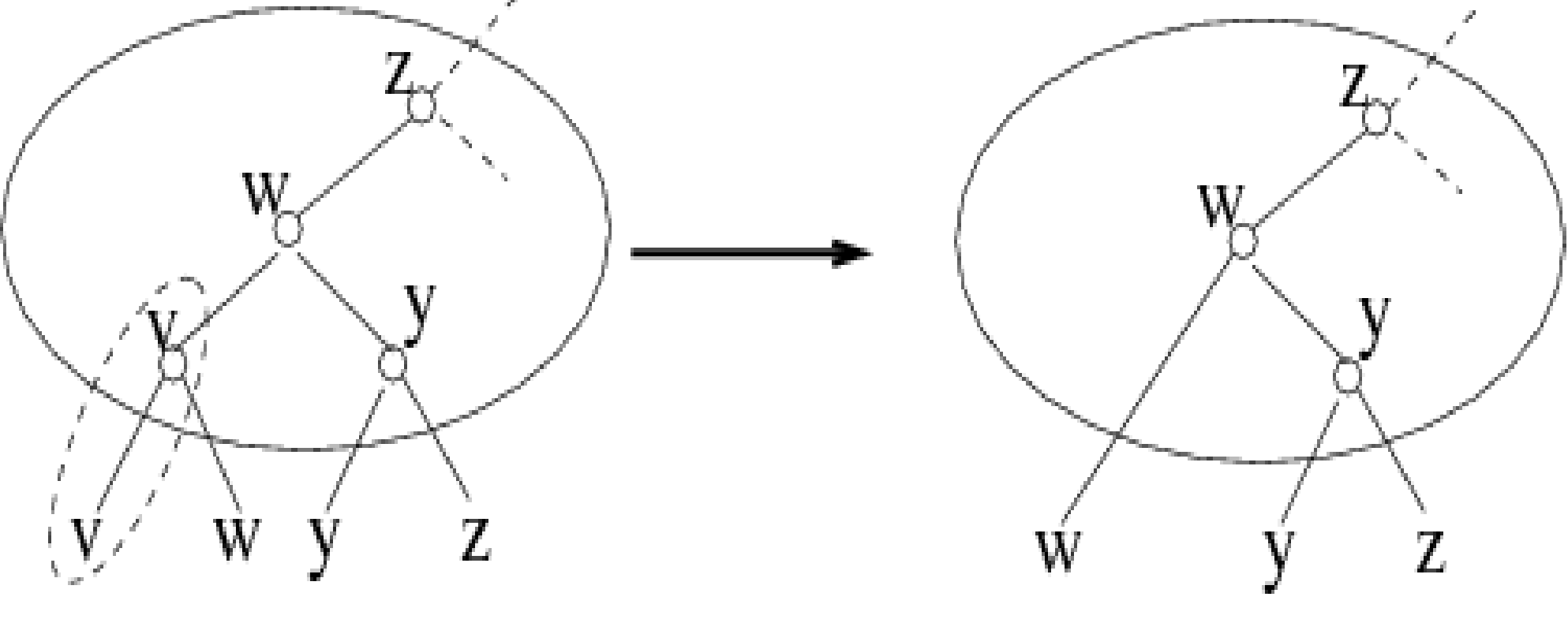}}\hspace{0.3in}
\subfigure[$w$ and $\helper(w)$ share a neighbor.]{\label{sfig: ldc2} \includegraphics[scale=0.3]{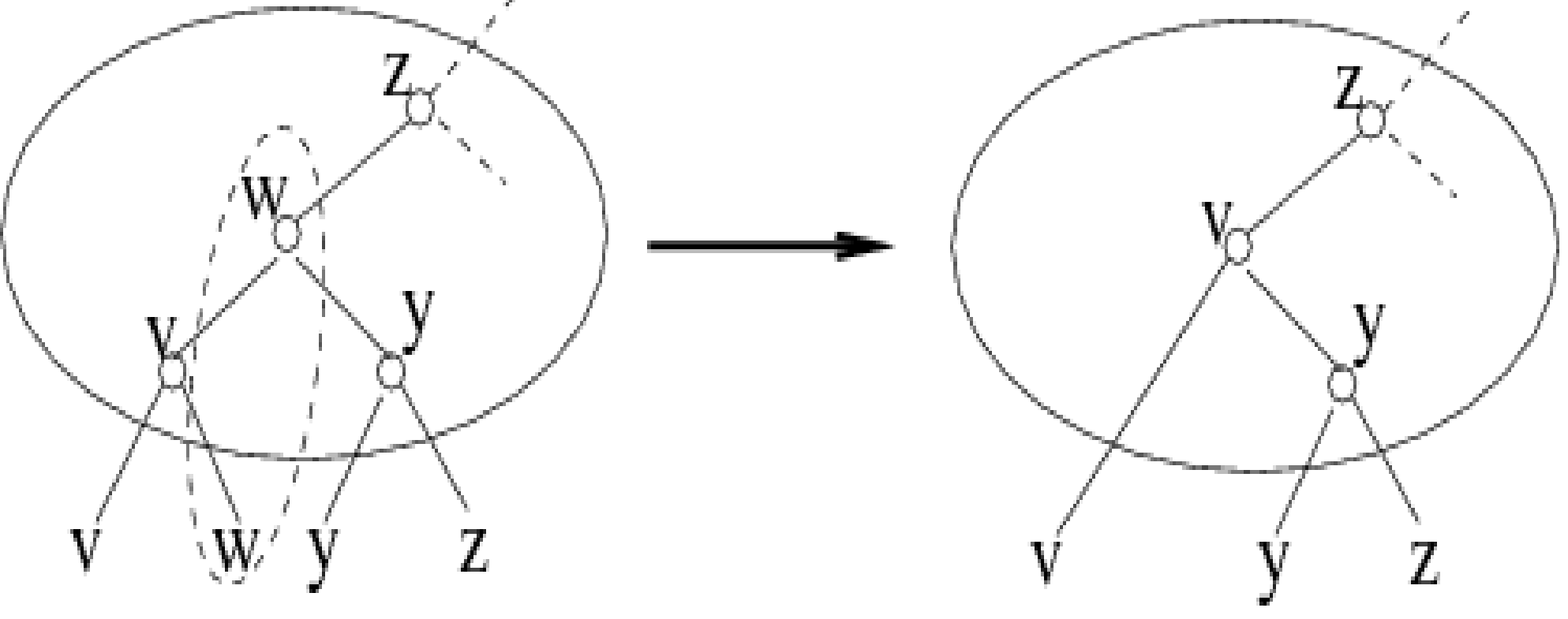}}\\
\subfigure[ $z$ and $\helper(z)$ do not share neighbors.]{\label{sfig: ldc3} \includegraphics[scale=0.3]{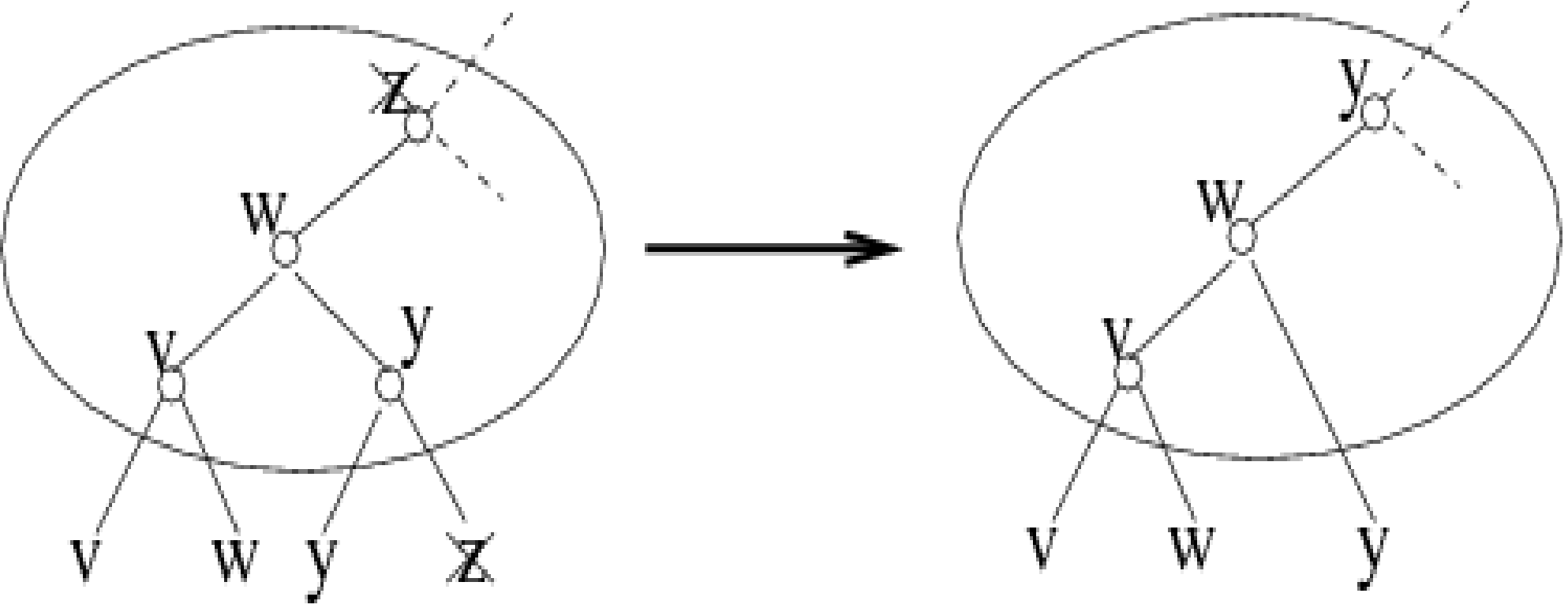}}\hspace{0.3in}
\subfigure[$z$ is an heir in Ready state.]{\label{sfig: lhdc0} \includegraphics[scale=0.3]{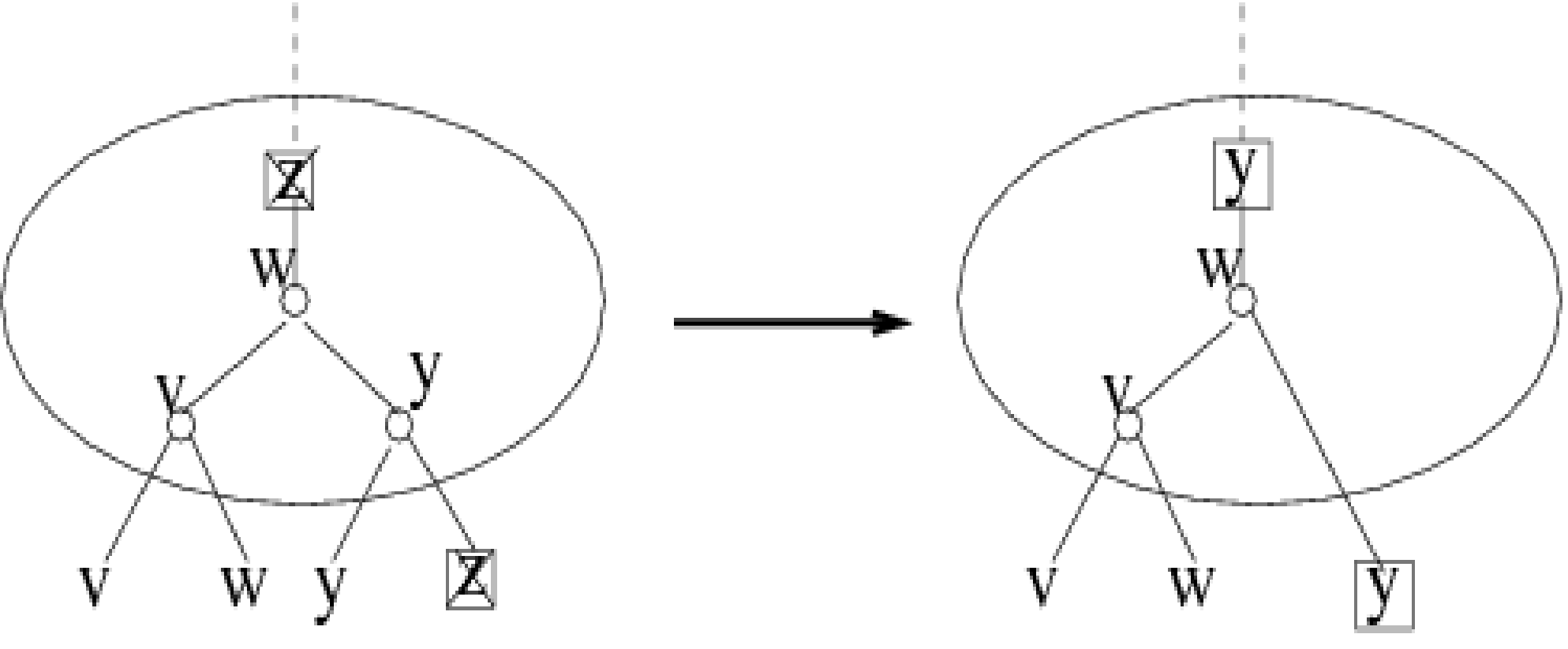}}
\caption{Various cases of Leaf deletions}
\end{figure}
\begin{figure}[t]
\centering
\includegraphics[scale=0.4]{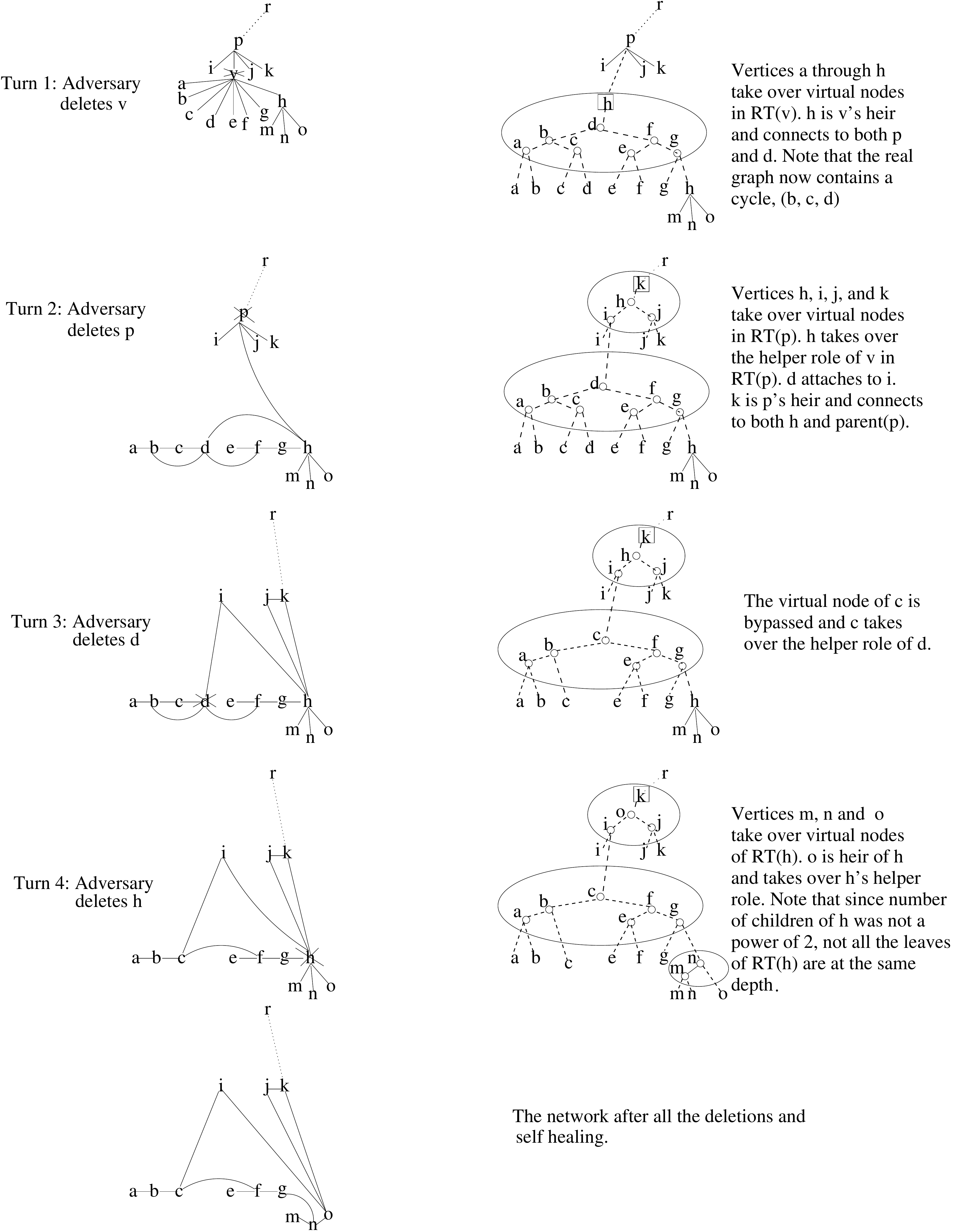}
\caption{An illustrative sequence of deletions and healings.}
\label{fig: story}
\end{figure}

 
 If the adversary removes a leaf node from the system, the healing is accomplished by its neighbors as specified in
Algorithm \ref{algo: fixleaf}: \textsc{FixLeafDeletion}. Let $v$ be the deleted leaf node and $p$ be its parent. Let us
consider the simple case first.  This is when the deleted node had no helper responsibility. This also implies its
original parent did not suffer a deletion. Node $p$ simply removes $v$ from the list of its children and then
recomputes and redistributes its will.\\
  Now, consider the situation where the deleted node had helper responsibilities.  In this case, the one node whose
workload has been reduced by this deletion is $p$. Using Algorithm \ref{algo: makeleafwill}:
\textsc{MakeLeafWill} $v$ hands over the list of  its helper responsibilities to $p$.  Here, a special
case may arise when $v$ is simulating a helper node which has $v$ itself as one of its $\hchildren$. Recall that $\parent(v)$ is $v$'s
ancestor closest to $v$ in the tree. This implies that $\parent(v) = \hparent(v) = p$. The only thing that $p$ needs
to do if $v$ is deleted is to remove $v$ from its $\hchildren$ and add itself (for consistency). This is the will
conveyed by $v$ to $p$. When $v$ is deleted, $p$ simply updates its helper fields.
 For other cases, $v$ simply sends its helper fields to $p$ to be copied to $p$'s helper reconstruction fields. In this
situation, when $v$ is actually deleted the following happens: the helper node that $p$ is simulating is deleted and
bypassed by the bypass operation defined earlier.  Node $p$ now simulates a new helper node that has the same helper
responsibilities previously fulfilled by $v$. In case the deleted leaf node was itself an heir in ready state, $p$
detects this and sets its flags accordingly. Again at the end of the reconstruction, the leaf nodes reconstruct their
wills. An example of such a leaf deletion is the deletion of  node $d$  at Turn 3 as shown in  figure \ref{fig: story}.

\par\noindent{\bf Important Note:}\ 
When implementing the pseudocode for Algorithm~\ref{algo: makewill}:
\textsc{MakeWill}, 
it is important to bear in mind that when $\RT(v)$ is being
updated due to a node deletion, most of $\SubRT(v)$ will be
unchanged.  In fact, only $O(1)$ nodes will need to have their
fields updated.  These can be found and updated more efficiently
by a more detailed algorithm based on case analysis, which we 
defer to the full version of this paper.

\floatname{algorithm}{Algorithm}

\begin{algorithm}[ph!]
\begin{algorithmic}[1]
\STATE Given a tree $T(V,E)$
\STATE \textsc{Init(T)}.
\WHILE {true}
\IF{a vertex $x$ is deleted}
\IF { $\children(x)$ is $\Empty$}
\STATE \textsc{FixLeafDeletion(x)}
\ELSE 
\STATE \textsc{FixNodeDeletion(x)}
\ENDIF
\ENDIF
\ENDWHILE
\end{algorithmic}
\caption{\textsc{Forgiving tree}: The main function.}
\label{algo: forgiving}
\end{algorithm}
 
 \begin{algorithm}[ph!]
\begin{algorithmic}[1]
\REQUIRE{each node of T has a unique ID}
\FOR{each node  $v\in T$ do} 
\STATE $\children(v)\leftarrow $ children of $v$.
\STATE $\parent(v)\leftarrow $ if $v$ is root of T then $\Empty$ else parent of $v$.
\STATE $\isreadyheir(v)\leftarrow false$. 
\STATE $\ishelper(v)\leftarrow false$.
\STATE $\hparent(v)\leftarrow \Empty$. 
\STATE $\hchildren(v)\leftarrow \Empty$. 
\STATE $\heir(v)\leftarrow$ if v is a leaf node then $\Empty$ else child of $v$ with highest ID.
\STATE $\SubRT(v)\leftarrow \textsc{generateSubRT}(v)$.
\STATE \textsc{MakeWill}($v, \SubRT(v)$).
\ENDFOR
\end{algorithmic}
\caption{\textsc{Init(T)}: initialization of the Tree T} 
\label{algo: init}
\end{algorithm}

\begin{algorithm}[ph!]
\caption{\textsc{FixNodeDeletion($v$)}: Self-healing on deletion of internal node }
\label{algo: fixnode}
\begin{algorithmic}[1]
\STATE \textsc{MakeRT}($\children(v),\parent(v)$).
\STATE \label{algline: ND2} let $h = \heir(v)$.Let $p = \parent(v)$
\IF {$\isreadyheir(h) =true$} 
\STATE  $\hparent(h)$ replaces $v$ by $h$ in $\SubRT(\hparent(h))$.
\STATE \textsc{MakeWill}($\hparent(h),\SubRT(\hparent(h))$).
\ENDIF \label{algline: ND6}
\FOR {each node $y\in \children(v)$}
 \IF {$\children(v)$ is $\Empty$}
\STATE \textsc{MakeLeafWill}($v$).
\ENDIF
\ENDFOR
\end{algorithmic}
\end{algorithm}

 \begin{algorithm}[ph!]
\caption{\textsc{FixLeafDeletion($v$)}: Self-healing on deletion of leaf node}
\label{algo: fixleaf}
\begin{algorithmic}[1]
\STATE let $p=\parent(v)$
\IF {$\ishelper(p)=false$}
\STATE $p$ removes $v$ from $\children(p)$
\STATE $SubRT(p)\leftarrow \textsc{GenerateSubRT}(p)$
\STATE \textsc{MakeWill}($p$)
\ELSE
\STATE Let $z=\parent(v)$.
\IF {$z \neq \hparent(v)$}
\STATE bypass($z$).
\ENDIF
\STATE $z $ makes edges with $\nexthparent(z)$,$\nexthchildren(z)$.
\STATE $\hparent(z)\leftarrow \nexthparent(z)$.
\STATE $\hchildren(z)\leftarrow \nexthchildren(z)$.
\IF{$\left|\hchildren(z)\right|$=1}
\STATE $\isreadyheir(z)=true $ 
\ENDIF
\ENDIF
\FOR {each node $y\in \children(v)$}
 \IF {$\children(v)$ is $\Empty$}
\STATE \textsc{MakeLeafWill}($v$).
\ENDIF
\ENDFOR
\end{algorithmic}
\end{algorithm}

  \begin{algorithm}[ph!]
\caption{\textsc{GenerateSubRT($\children(v)$):} Computes  the Reconstruction Tree ($\RT$) of $v$ minus a possible helper node
simulated by $\heir(v)$. }
\label{algo: genRT}
\begin{algorithmic}[1]
\STATE Let $Lset$ be a set of vertices representing all members of $\children(v)$, and $Iset$ be another set of vertices.
representing  all members of $\children(v)$ except the one with the highest $ID$.
\STATE Arrange $Lset$ in ascending order of their $ID$s.
\STATE Using the arranged $Lset$ as leaves and $Iset$ as the internal nodes construct a  Balanced Binary Search Tree
$\SubRT$  ordered on the nodes $ID$.
\RETURN $\SubRT$
\end{algorithmic}
\end{algorithm}

  \begin{algorithm}[ph!]
\caption{\textsc{MakeWill($v,\SubRT(v))$:} Makes and distributes the will of v }
\label{algo: makewill}
\begin{algorithmic}[1]
\STATE Let p = $\parent(v)$. Let  $rv$ be root of $\SubRT(v)$.
\FOR {each node $y\in \children(v)$}
    \STATE let $ly$ be the leaf vertex representing $y$ in $\SubRT(v)$. Let $hy$ be the internal node in $\SubRT$ representing $y$. 
    \STATE If $hy$ is $ly$'s parent in $\SubRT(v)$ then $\nextparent(y)\leftarrow$ parent of $hy$ in $\SubRT$ else $\nextparent(y)\leftarrow$ parent of $ly$ in $\SubRT$.
     \IF{ $y\neq heir(v)$}
     \STATE $\nexthchildren(y)\leftarrow$ children of $hy$ in $\SubRT$.   
     \STATE $\nexthparent(y)\leftarrow$ parent of $hy$ in $\SubRT$.  
       \ELSE
        \IF{$\ishelper(v) = true$}
          \STATE $\nexthchildren(y)\leftarrow \hchildren(v)$.
          \STATE $\nexthparent(y)\leftarrow \hparent(v)$.
          \STATE $\nexthparent(rv)\leftarrow p$.     
         \ELSE
                 \STATE $\nexthchildren(y)\leftarrow rv$.
                  \STATE $\nexthparent(y)\leftarrow p$.
                  \STATE $\nexthparent(rv)\leftarrow y$.
         \ENDIF
     \ENDIF
\ENDFOR
\end{algorithmic}
\end{algorithm}

\begin{algorithm}[ph!]
\caption{\textsc{MakeLeafWill($v$)}:  Leaf node leaves a will for its parent.}
\label{algo: makeleafwill}
\begin{algorithmic}[1]
\STATE let $z=\parent(v)$.
\IF{$z = \hparent(v)$}
\STATE $\nexthparent(z)\leftarrow \hparent(v)$.
 \STATE $\nexthchildren(z)\leftarrow \hchildren(z)/\{v\} \cup \{z\}$. \COMMENT{$z$ will take on itself as a child of
its helper node.}
 \ELSE
\STATE $\nexthparent(z)\leftarrow \hparent(v)$.
\STATE $\nexthchildren(z)\leftarrow \hchildren(v)$.
\ENDIF
\end{algorithmic}
\end{algorithm}

\begin{algorithm}[ph!]
\caption{\textsc{makeRT(children(v),parent(v))}: Replace the deleted node by its  $\RT$}
\label{algo: makeRT}
\begin{algorithmic}[1]
\FOR{each node $x\in$ $\children$(v)}
  \IF{$\isreadyheir(x)=true$}
  \STATE bypass($x$).     \COMMENT{ $\hparent(x)$ and $\hchildren(x)$ bypass $x$ and connect themselves. }
       \STATE \textsc{MakeHelper}($x$).
  \ELSE     
\STATE $x$ makes edge between itself and $\nextparent(x)$.
\STATE $\parent(x)\leftarrow \nextparent(x)$.
\STATE \textsc{MakeHelper}($x$).
\ENDIF
\ENDFOR 
\end{algorithmic}
\end{algorithm}

\begin{algorithm}[h!]
\caption{\textsc{MakeHelper($v$)}: $v$ takes over helper node responsibilities}
\label{algo: makehelper}
\begin{algorithmic}[1]
\STATE $v$ makes edges between itself and $\nexthchildren(v)$, and $\nexthparent(v)$.
\STATE $\hparent(v)\leftarrow \nexthparent(v)$.
\STATE $\hchildren(v)\leftarrow \nexthchildren(v)$.
\STATE $\ishelper(v) = true$.
\IF{$|\hchildren(v)| = 1$}
 \STATE $\isreadyheir(v) = true$.  \COMMENT{Only an 'unemployed' $\heir$ has a single child.}
 \ENDIF
\end{algorithmic}
\end{algorithm}

\section{Results}
\label{subsec: Results}

\subsection{Upper Bounds}
\label{subsec: upperbounds}

Let $T$ be  the original tree, let $D$ be its diameter, and let $\Delta$ be its maximum degree.

\begin{theorem}
The Forgiving Tree has the following properties:
\label{theorem: forgiving}
\begin{enumerate}
\item\label{th: degree} 
  The Forgiving Tree increases the degree of any vertex by at most $3$.
\item\label{th: diameter} 
  The Forgiving Tree always has diameter $O(D \log \Delta)$.
\item \label{th:cost} 
  The latency per deletion and number of messages sent per node per 
  deletion is $O(1)$; each message contains $O(1)$ bits and node IDs.
\end{enumerate}
\end{theorem}
\begin{proof}
 Parts~\ref{th: degree} and~\ref{th:cost} follow directly by
construction of our algorithm.  For part~\ref{th: degree}, we note
that for a node $v$, any degree increase for $v$ is imposed by its
edges to $\hparent$($v$) and $\hchildren(v)$. By construction, node
$v$ can play the role of at most one helper node at any time and the
number of $\hchildren$ is never more than $2$, because the
reconstruction trees are binary trees.  Thus the total degree increase
is at most $3$.  Part~\ref{th:cost} also follows directly by the
construction of our algorithm, noting that, because the virtual nodes all have degree at most $3$, healing one deletion 
results in at most $O(1)$ changes to the edges in each 
affected reconstruction tree.  In fact, the changes to 
$\RT(w)$ for an affected node $w$ do not require new information,
which allows these messages to be computed and distributed in parallel.
We leave the details to the full version due to space constraints.

We next show Part~\ref{th: diameter}, that the diameter of the
Forgiving Tree is always $O(D \log \Delta)$.  We assume that $T$ is
rooted with some root $r$ selected arbitrarily and let $p(v)$ be the
parent of $v$ in $T$.  Fix any node $v_{1}$ in the tree and let $P =
v_1,\ldots ,v_{\ell}$ be the path in $T$ from the node $v_{1}$ to the
root $v_{\ell} = r$.  Let $T'$ be the Forgiving Tree data structure
after some fixed number of deletions.  To simplify notation, if a node
$v$ has not been deleted in $T'$, we will let root of $\RT(v)$ be $v$
by default.  We construct a path $P'$ through $T'$ from $v_{1}$ to the
root of $RT(r)$ as follows.  For all $i$ from $1$ to $\ell$, $P'$
connects the root of $\RT(v_{i})$ to the root of $\RT(v_{i+j})$ for
the smallest $j \geq 1$ such that $\RT(v_{i+j})$ exists.  We note that
if any nodes remain in $T'$, then $\RT(r)$ must still exist and so
$P'$ connects $v_{1}$ to the root of $\RT(r)$.  We further note that
the length of $P'$ is no more than a factor of $\log \Delta + 1$
greater than the length of $P$.  To see this, consider the length of
the path connecting the root of $\RT(v_{i})$ to the root of
$\RT(v_{i+j})$ for the smallest $j \geq 1$ such that $\RT(v_{i+j})$
exists.  Such a path can have length at most equal to the log of the
number of children of $v_{i+j}$ in $T$, a quantity that can be at most
$\log \Delta + 1$.  The length of the path between the roots will not
increase even when nodes in $\RT(v_{i+j})$ are deleted.

Now consider any pair of nodes $x$ and $y$ in $T'$.  We know from the
above that $x$ and $y$ are connected to the root of $\RT(r)$ in $T'$
by paths of length at most $h \cdot (\log \Delta + 1)$, where $h$ is
the height of $T$.  We further know that the diameter of $T$, $D$, is
at least $h$.  It follows directly that the diameter of the Forgiving
Tree is never more than $O(D \log \Delta)$.
\end{proof}

\subsection{Lower Bounds}
\label{subsec: lowerbounds}

\begin{theorem}
Consider any self-healing algorithm that ensures that: 1) each node
increases its degree by at most $\alpha$, for some $\alpha \geq 3$;
and 2) the diameter of the graph increases by a multiplicative factor
of at most $\beta$.  Then for any positive $\Delta$, for some initial
graph with maximum degree $\Delta$, it must be the case that
$\alpha^{2\beta + 1} \geq \Delta$.
\end{theorem}

\begin{proof}
Let $G$ be a star on $\Delta + 1$ vertices, where $x$ is the root
node, and $x$ has $\Delta$ edges with each of the other nodes in the
graph.  Let $G'$ be the graph created after the adversary deletes the
node $x$.  Consider a breadth first search tree, $T$, rooted at some
arbitrary node $y$ in $G'$.  We know that the self-healing algorithm
can increase the degree of each node by at most $\alpha$, thus every
node in $T$ can have at most $\alpha+1$ children.  Let $h$ be the
height of $T$.  Then we know that $1 + (\alpha + 1) \sum_{i=0}^{h-1}
\alpha^{i} \geq \Delta$.  This implies that $\alpha^{h+1} \geq \Delta$
for $\alpha \geq 3$, or ${h+1} \geq \log_{\alpha} \Delta$.  Since the
diameter of $G$ is $2$, we know that $\beta = h/2$, and thus $2\beta +
1 \geq \log_{\alpha} \Delta$.  Raising $\alpha$ to the power of both
sides, we get that $\alpha^{2\beta + 1} \geq \Delta$.
\end{proof}

\medskip
\noindent
We note that this lower-bound compares favorable with the general
result achieved with our data structure.  The Forgiving Tree can be
modified so that it ensures that 1) the degree of any node increases
by no more than $\alpha$ for any $\alpha \geq 3$; and that the
diameter increases by no more than a multiplicative factor of $\beta
\leq 2 \log_{\alpha} \Delta + 2$.  Thus we can ensure that
$\alpha^{(1/2)(\beta - 1)} \leq \Delta$.

\section{Conclusion}

We have presented a distributed data structure that withstands
repeated adversarial node deletions by adding a small number of new
edges after each deletion.  Our data structure ensures two key
properties, even when up to all nodes in the network have been
deleted.  First, the diameter of the network never increases by more
than $O(\log \Delta)$ times its original diameter, where $\Delta$ is
the maximum original degree of any node.  For many peer-to-peer
systems, $\Delta$ is at most polylogarithmic, and so the diameter
would increase by no more than a $O(\log \log n)$ multiplicative
factor.  Second, no node ever increases its degree by more than $3$
over its original degree.

Several open problems remain including the following.  Can we protect
other invariants?  For example, can we design a distributed data
structure to ensure that the stretch between any pair of nodes
increases by no more than a certain amount?  Can we design algorithms
for less flexible networks such as sensor networks?  For example, what
if the only edges we can add are those that span a small distance in
the original network?  Finally, can we extend the concept of
self-healing to other objects besides graphs?  For example, can we
design algorithms to rewire a circuit so that it maintains its
functionality even when multiple gates fail?


\bibliography{selfheal} 
\bibliographystyle{latex8}

\end{document}